\documentclass[11pt]{article}
\usepackage{amsfonts,epsfig,latexsym,amscd}

\newcommand{\R}{{\mathbb{R}}}
\newcommand{\Z}{{\mathbb{Z}}}

\newcommand{\C}{{\mathbb{C}}}
\newcommand{\I}{{\mathbb{I}}}
\newcommand{\CP}{{\mathbb{C}}{{P}}}
\newcommand{\RP}{{\mathbb{R}{{P}}}}
\newcommand{\beq}{\begin{equation}}
\newcommand{\eeq}{\end{equation}}
\newcommand{\bea}{\begin{eqnarray}}
\newcommand{\eea}{\end{eqnarray}}
\newcommand{\ra}{\rightarrow}

\newcommand{\cd}{\partial}
\newcommand{\wt}{\widetilde}
\newcommand{\wh}{\widehat}
\newcommand{\ol}{\overline}
\newcommand{\sigvec}{\mbox{\boldmath{$\sigma$}}}
\newcommand{\lamvec}{\mbox{\boldmath{$\lambda$}}}
\newcommand{\tauvec}{\mbox{\boldmath{$\tau$}}}

\newcommand{\nvec}{\mbox{\boldmath{$\wh{n}$}}}

\newcommand{\Zvec}{{\bf Z}}

\newcommand{\msn}{{\mathfrak M}_n}
\newcommand{\msnt}{\wt{\mathfrak M}_n}
\newcommand{\msit}{\wt{\mathfrak M}_1}
\newcommand{\gamt}{\wt{\gamma}}
\newcommand{\msi}{{\mathfrak M}_1}

\newcommand{\vol}{{\rm vol}}
\newcommand{\holo}{\mathsf{Hol}}

\newcommand{\Rrot}{{\cal R}}

\newcommand{\Lrot}{{\cal L}}
\newcommand{\Pp}{\mathsf{P}}

\newcommand{\tr}{{\rm tr}}

\newtheorem{thm}{Theorem}

\newcommand{\id}{{\rm Id}}

\begin{document}

\title{The deformed conifold as a geometry on the space of unit charge
$\CP^1$ lumps \\[30pt]}
\author{J.M. Speight\\[10pt]
{\sl Department of Pure Mathematics, University of Leeds} \\
{\sl Leeds LS2 9JT, England}\\
{\sl E-mail: j.m.speight@leeds.ac.uk}}

\date{}

\maketitle

\begin{abstract}

The strong structural similarity between the deformed conifold of
Candelas and de la Ossa (a noncompact Calabi-Yau manifold) and the moduli 
space of unit charge $\CP^1$ lumps equipped with its $L^2$ metric is
pointed out. This allows one to reinterpret certain recent results on
D3 branes in terms of lump dynamics, and to deduce certain curvature
properties of the deformed conifold.

\end{abstract}

\section{Introduction}
\label{intro}

The purpose of this note is to point out that the deformed conifold of
Candelas and de la Ossa \cite{candel} and the $L^2$ geometry on the 1-lump
moduli space  $\msi$
 of the $\CP^1$ model both lie in a certain class of
cohomogeneity 1 K\"ahler geometries on $PL(2,\C)$. Consequently, they have 
strong structural similarities which may be used to transfer results obtained
for one to the other, often with a complete change in physical 
interpretation. For example, their Hodge isomorphisms restricted to 
cohomogeneous 3-forms will be shown to coincide. This allows one to 
transfer a family of harmonic 3-forms constructed by Cvetic et al
\cite{cvegib} from the deformed conifold to $\msi$ changing their interpretation
from D3 brane solutions of supergravity to supersymmetric quantum lump
states. Thinking of the deformed conifold as a geometry on $\msi$ also gives
an elementary proof that it is the unique scalar flat geometry in its
class.

\section{The deformed conifold and $\CP^1$ lumps}
\label{defor}

The deformed conifold is $TS^3$ equipped with its unique $SO(4)$ invariant
Ricci-flat K\"ahler metric, $\gamt$. Thinking of $TS^3$ embedded in
$\R^4\oplus\R^4$ as $\{(X,\xi):|X|=1,X\cdot\xi=0\}$, the isometric $SO(4)$
action is simply
\beq
{\cal O}:(X,\xi)\mapsto ({\cal O}X,{\cal O}\xi).
\eeq
This action has cohomogeneity 1, the orbits being level sets of $|\xi|\in
[0,\infty)$. It may be viewed as a ``cone'' with link $S^2\times S^3$, but
with the singular tip at $\xi=0$ replaced by $S^3$, the single exceptional 
orbit of  the $SO(4)$ action. The complex structure with respect to which
$\gamt$ is K\"ahler is less obvious; it is obtained by identifying
$TS^3$ with $SL(2,\C)$ via
\beq
M:=Z_0\I_2+i\Zvec\cdot\tauvec\in SL(2,\C),\qquad
Z:=\cosh|\zeta|\, X+i\frac{\sinh|\zeta|}{|\zeta|}\zeta\in\C^4,
\eeq
where
$\tauvec=(\tau_1,\tau_2,\tau_3)$ are the Pauli spin matrices.
Using the isomorphism ${\rm Spin}(4)\cong SU(2)\times SU(2)$, one
may equally well think of $\gamt$ as a $SU(2)\times SU(2)$ invariant 
K\"ahler metric on $SL(2,\C)$, the isometric action here being
\beq
(L,R):M\mapsto LMR.
\eeq
Then $(-\I_2,\I_2)$ generates a $\Z_2$ subgroup acting isometrically,
holomorphically and properly discontinuously on $SL(2,\C)$ so that $\gamt$
descends to a $PU(2)\times PU(2)$ invariant K\"ahler metric $\gamma$ on
$PL(2,\C)=SL(2,\C)/\Z_2$. Conversely $\gamma$ uniquely defines $\gamt$ by
lifting.

Now each $[M]=\{\pm M\}\in PL(2,\C)$ may be uniquely identified with a 
static $\CP^1$ lump on $S^2$, that is, a degree 1 holomorphic map
$\phi:S^2\ra\CP^1\cong S^2$. Explicitly, one identifies
\beq
\label{rat}
\pm\left(\begin{array}{cc}
a & b \\ c & d \end{array}\right)
\qquad\leftrightarrow\qquad
W:z\mapsto\frac{az+b}{cz+d}
\eeq
where $z,W\in\C$ are stereographic coordinates on the domain and codomain.
Map composition and projective matrix multiplication coincide under this
identification. Further, the left and right $PU(2)$ multiplication actions
on $PL(2,\C)$ coincide with the SO(3) rotation actions on the codomain and 
domain respectively. Hence $\gamma$ may naturally be interpreted as a
$SO(3)\times SO(3)$ invariant K\"ahler metric on $\msi$, the 1 lump
moduli space. This class of metrics has
 been studied in depth because $\gamma_{L^2}$,
the $L^2$ metric on $\msi$ (the restriction to $T\msi$ of the $\CP^1$ model
kinetic energy functional), belongs to it 
\cite{spe1,spe2}. Recall that geodesics in $(\msi,\gamma_{L^2})$ are 
believed to be close to low-energy dynamical solutions of the $\CP^1$
model, in analogy with the method of Manton \cite{war,man}.

The most natural coordinate system in which to analyze these metrics is 
obtained by projecting the standard polar decomposition of  $SL(2,\C)$ to
$PL(2,\C)$. That is, each $\{\pm M\}\in PL(2,\C)$ is uniquely decomposed
as
\beq
M=U(\Lambda\I_2+\lamvec\cdot\tauvec)
\eeq
where $\pm U\in SU(2)$, $\lamvec\in\R^3$ and
$\Lambda:=\sqrt{1+\lambda^2}$. Physically
one interprets the map with decomposition $([U],\lamvec)$ as a lump located
at $-\lamvec/\lambda\in S^2$ with sharpness $\lambda$ and internal
orientation $[U]\in PU(2)\cong SO(3)$ \cite{spe1}. If $\lambda=0$ the
map has constant energy density. So these coordinates give a very natural
parametrization of $\msi$ from the physical viewpoint. The $SO(3)\times
SO(3)$ action is particularly simple in these terms also:
\beq
(\Lrot,\Rrot):([U],\lamvec)\mapsto([LUR],\Rrot\lamvec)
\eeq
where $[L]\in PU(2)$ is related to $\Lrot\in SO(3)$ by the 
canonical isomorphism
\beq
\Lrot_{ab}=\frac{1}{2}\tr(\tau_aL^\dagger\tau_bL),
\eeq
and similarly for $[R]$ and $\Rrot$. One sees once again that the action has
cohomogeneity 1, the orbits being level sets of $\lambda\in[0,\infty)$.
Every orbit except $\lambda=0$ is diffeomorphic to $S^2\times SO(3)$, while
$\lambda=0$ is diffeomorphic to $SO(3)$.

Denoting the usual left-invariant 1-forms on $PU(2)$ by $\sigvec=(\sigma_1,
\sigma_2,\sigma_3)$, the pulled back action on the moving coframe
$(\sigvec,d\lamvec)$ is
\beq
(\Lrot,\Rrot)^*:(\sigvec,d\lamvec)\mapsto(\Rrot\sigvec,\Rrot d\lamvec).
\eeq
So in this coordinate system the problem of constructing 
$SO(3)\times SO(3)$ invariant
$(0,p)$ tensors on $M_1$ is reduced to constructing $SO(3)$ scalars of
rank $p$ from the 1-form triplets $\sigvec$, $d\lamvec$. Applying this idea
to any invariant metric, its K\"ahler form and Ricci tensor, it is possible 
to prove the following \cite{spe2}:

\begin{thm}
\label{isomprop} Let $\hat{\gamma}$ be a $SO(3)\times SO(3)$ invariant 
K\"ahler metric on $\msi$ and $\rho$ be its Ricci tensor. Then there
exists a smooth function $A:[0,\infty)\ra\R$ such that
\beq
\label{tensor}
\hat{\gamma}=
A_1d\lamvec\cdot d\lamvec +A_2(\lamvec\cdot d\lamvec)^2+
A_3\sigvec\cdot\sigvec
+A_4(\lamvec\cdot\sigvec)^2
+A_1\lamvec\cdot(\sigvec\times d\lamvec),
\eeq
where $A_1,\ldots,A_4$ are functions of $\lambda$ only, determined by
$A$ as
\beq
A_1=A(\lambda),\quad
A_2=\frac{A(\lambda)}{\Lambda^2}+\frac{A'(\lambda)}{\lambda},\quad
A_3=\left(\frac{\Lambda^2+\lambda^2}{4}\right)A(\lambda),\quad
\label{Aconstraints}
A_4=\frac{\Lambda^2}{4\lambda}A'(\lambda).
\eeq
Further, the Ricci tensor takes the same form, namely,
\beq
\rho=
\bar{A}_1d\lamvec\cdot d\lamvec +\bar{A}_2(\lamvec\cdot d\lamvec)^2+
\bar{A}_3\sigvec\cdot\sigvec
+\bar{A}_4(\lamvec\cdot\sigvec)^2
+\bar{A}_1\lamvec\cdot(\sigvec\times d\lamvec),
\eeq
$\bar{A}_1,\ldots,\bar{A}_4$ being functions of $\lambda$ only, 
determined by a single function
$\bar{A}$ as in (\ref{Aconstraints}), where
\beq
\label{Abardef}
\bar{A}=
-\frac {2\lambda\Lambda^2(A')^2+
(9\lambda^2+4)AA'+\lambda\Lambda^2AA''+4\lambda A^2 }
{2\lambda A[(\Lambda^2+\lambda^2)A+\lambda\Lambda A']}.
\eeq
\end{thm}

In the case of the deformed conifold, $\gamma$ is Ricci flat and hence
$\bar{A}\equiv 0$. This is a second order ODE for $A(\lambda)$ which is
easily reduced  to quadratures. The unique solution which is regular on the
exceptional orbit $\lambda=0$ (i.e.\ with $A'(0)=0$) is
\beq
\label{flat}
A(\lambda)=e^{\int_0^\lambda\beta(\mu)d\mu},\qquad
\beta(\lambda)=-\frac{(4\lambda^5+4\lambda^3+3\lambda)\Lambda-
(6\lambda^2+3)\sinh^{-1}\lambda}
{3\lambda[(2\lambda^5+3\lambda^2+\lambda)\Lambda-
(\lambda^2+1)\sinh^{-1}\lambda]}.
\eeq
This should be compared with the $L^2$ metric on $\msi$,
\beq
A_{L^2}=\frac{\mu(\mu^4-4\mu^2\log\mu-1)}{(\mu^2-1)^3},\quad
\mu:=(\Lambda+\lambda)^2.
\eeq
The Ricci flat
component function (\ref{flat}) does look somewhat more complicated, 
but in practice it is often the structure of $\gamma$ (\ref{tensor}),
(\ref{Aconstraints})
which is important rather than the details of $A(\lambda)$.  
Using the elementary estimate $A(\lambda)>\lambda^{-5/6}$ for all $\lambda$
sufficiently large, it is 
straightforward to verify directly the well-known geometric properties of
the deformed conifold (infinite volume and diameter, geodesic completeness).

Much insight
into $\gamma$ may be obtained from $\gamma_{L^2}$ and {\it vice versa}.
We shall see in the next section that a certain family of harmonic
forms on $(\msi,\gamma)$ transfers directly to $(\msi,\gamma_{L^2})$, 
providing a link bewteen D3 branes and quantum lump dynamics.
Other results transfer the other way. For example the explicit formulae 
(in terms
of $A$) for holomorphic sectional curvatures, and the analysis of
cohomogeneity 1 Hamiltonian flows on $\msi$ obtained for
$\gamma_{L^2}$ in  \cite{spe2} apply equally well to the deformed confifold.

\section{Scalar flatness}
\label{scala}

One
may prove a stronger uniqueness
 result for the deformed conifold,
namely

\begin{thm} The deformed conifold is the unique 
$SO(4)$ invariant scalar flat K\"ahler metric on $TS^3$.
\end{thm}

\noindent
Since $\kappa=0$ does {\em not} imply $\rho=0$ (rather merely $\tr\, \rho=0$)
this is a stronger statement than we had 
previously (unique $SO(4)$ invariant {\em Ricci} flat K\"ahler metric). 
As shown above, it suffices to prove that $\gamma$ is the unique
$SO(3)\times SO(3)$ invariant K\"ahler solution of $\kappa=0$. 
From theorem \ref{isomprop} it
immediately follows for any invariant K\"ahler metric on $\msi$ that
\beq
\label{24}
\kappa
=4\frac{\bar{A}}{A}+2\frac{\bar{B}}{B}.
\eeq
where
\beq
\label{Bdef}
B=\frac{1}{4}[(\Lambda^2+\lambda^2)A+\lambda\Lambda A']
\eeq
and $\bar{B}$ is similarly defined in terms of $\bar{A}$ \cite{spe2}. 
Substituting (\ref{Abardef}) and (\ref{Bdef}) into 
the scalar flat condition, $\kappa=0$, it reduces to a single
third order nonlinear ODE for $A$, to which we seek nonsingular,
strictly positive solutions.

 Assume there exists a point at which $\bar{A}
\neq 0$. Then by continuity, there exists an interval on which $\bar{A}\neq
0$. On this interval, we may rearrange $\kappa=0$, using (\ref{24}), so that
\bea
\frac{\bar{B}}{\bar{A}}&=&-2\frac{B}{A} \nonumber \\
\Rightarrow
\frac{1+2\lambda^2}{4}+\frac{\lambda+\lambda^3}{4}\frac{\bar{A}'}{\bar{A}}
&=&-2\left(
\frac{1+2\lambda^2}{4}+\frac{\lambda+\lambda^3}{4}\frac{A'}{A}\right)
\nonumber \\
\Rightarrow
\frac{d\, }{d\lambda}[\log(A^2\bar{A})]&=&-3\frac{(1+2\lambda^2)}{\lambda+
\lambda^3} \nonumber \\
\label{25}
\Rightarrow
A^2\bar{A}&=&\frac{C}{(\lambda\Lambda)^3}
\eea
where $C>0$ is a constant. Now $\bar{A}$ cannot vanish at the ends of the
interval, or else $A$ is singular, by (\ref{25}). Hence, the interval 
must be $(0,\infty)$, that is, if $\bar{A}\neq 0$ anywhere, then
$\bar{A}$ vanishes nowhere. But then (\ref{25}) holds globally, so $A$
or $\bar{A}$ or both must be singular at $\lambda=0$, a contradiction.
Hence, all regular solutions of $\kappa=0$ have $\bar{A}=0$ everywhere, and
are thus Ricci flat. The metric $\gamma$ constructed above  
being the unique such metric, the theorem is proved. $\Box$

\section{Harmonic 3-forms}
\label{harmo}

The original motivation behind the construction of $(TS^3,\gamt)$ was to 
find a nontrivial Calabi-Yau manifold to be used as the transverse space to
Minkowski space $(\R^4,\eta)$ in superstring theory \cite{candel}. It has 
excited renewed interest recently due to the observation that certain
warped products $(\R^4\times TS^3,\alpha\eta\oplus\alpha^{-1}\gamt)$, where
$\alpha:TS^3\ra\R$, have the interpretation of D3-brane solutions of
supergravity \cite{brane}. 
The key step in constructing such a solution is to find a 
(complex) self dual harmonic 3-form on $(TS^3,\gamt)$; this in turn
determines $\alpha$. Those forms of pure holomorphicity type are
interpreted as supersymmetric solutions. 

It turns out that the Hodge isomorphism $*:\bigwedge^3\msi\ra\bigwedge^3\msi$
restricted to $SO(3)\times SO(3)$ invariant 3-forms is independent of $A$.
This is somewhat surprising: although the Hodge automorphism on middle
rank forms in even dimensions is always {\em conformally} 
invariant, the different
metrics introduced in theorem \ref{isomprop} are not necessarily conformally
related. In fact, assuming a normalization such as $A(0)\equiv 1$, no two
metrics are conformally equivalent. The invariance of $*$ stems from the
block structure of $\hat{\gamma}$ and the way in which the space of
$SO(3)\times SO(3)$ invariant 3-forms sits in $\bigwedge^3\msi$. It certainly
does not extend to general 3-forms. In any case,
an invariant 3-form is harmonic with respect to $\gamma$ if and only
if it is harmonic with respect to $\gamma_{L^2}$. Cvetic et al, as part of
a much bigger analysis of Stenzel metrics (Ricci-flat K\"ahler metrics on
$TS^n$, $n=2,3,\ldots$),  have
constructed a 4-dimensional family of such forms (although they do not
mention $SO(4)$ invariance) which transfers directly to 
$(\msi,\gamma_{L^2})$. How might one interpret them in this new setting?

The standard quantum mechanics of Bogomol'nyi solitons is determined
\cite{gibman} by a covariant Schr\"odinger equation for a wavefunction
$\psi$ on the soliton moduli space, in this context $\psi:\msi\times \R\ra
\C$ where
\beq
i\frac{\cd\psi}{\cd t}=-\frac{1}{2}\Delta_{\gamma_{L^2}}\psi.
\eeq
Gauntlett \cite{gau} has proposed, in the case of $\sigma$-models with
K\"ahler target space, a supersymmetric generalization 
of this in which $\psi$
is an antiholomorphic $p$-form, where
$p=0,\ldots,{\rm dim}_\C\msi=3$, and $\Delta_{\gamma_{L^2}}$ is replaced
by the Laplacian on forms $dd^\dagger+d^\dagger d$ (here
$d^\dagger:=*d*$). Hence antiholomorphic {\em harmonic} 3-forms have an
alternative interpretation as supersymmetric quantum lump states which
saturate the classical Polyakov energy bound, with the maximal number of
fermionic zero modes excited.

The easiest way to translate the relevant results of \cite{cvegib} into
$\CP^1$ model language is to rederive them, emphasising the $SO(3)\times
SO(3)$ invariance property. This will also allow 
us to strengthen the results 
slightly: we will show that the 4-dimensional space we construct exhausts
the invariant 3-forms which are closed and co-closed. 

Let $\psi$ be a $SO(3)\times SO(3)$ invariant 3-form on $\msi$. Then $\psi$
is uniquely determined by specifying its value at one point on each orbit,
for example at $M_\lambda=\Lambda\I_2+\lambda\tau_3$, 
$\lambda\geq 0$. In other
words, $\psi$ is uniquely determined by the 1-parameter family of
alternating trilinear forms $\psi_\lambda:V_\lambda\oplus V_\lambda\oplus
V_\lambda\ra\R$, where $V_\lambda:=T_{M_\lambda}\msi$. Further, each 
$\psi_\lambda$ must be invariant under the isotropy group
$H_\lambda<SO(3)\times SO(3)$ of $M_\lambda$, explicitly,
\beq
\label{26}
H_\lambda=\left\{\begin{array}{ll}
\{(\pm\exp(-\frac{i}{2}\theta\tau_3),\pm\exp(\frac{i}{2}\theta\tau_3)):
\theta\in\R\}
\cong SO(2) & \lambda>0 \\
\{(\pm U^\dagger,\pm U):\{\pm U\}\in SU(2)/\Z_2\}\cong SO(3) & \lambda=0.
\end{array}\right.
\eeq
Now the character of the induced representation of $H_\lambda$ on
$\bigwedge^3V_\lambda^*$ is
\beq
\label{28}
\chi(\theta)=8+8\cos\theta+4\cos 2\theta.
\eeq
Note that this formula holds for all $\lambda$; if $\lambda=0$ it is to be
interpreted as a function on $SO(3)$ independent of the axis of rotation
$\nvec\in S^2$ and depending only on the angle of rotation $\theta$.
Taking the character inner product of $\chi$ and the trivial character
$\chi_0(\theta)=1$ with respect to the Haar measure $d\mu_\lambda$ on
$H_\lambda$ ($d\mu_\lambda=(2\pi)^{-1}d\theta$, $\lambda>0$;
$d\mu_0={\pi}^{-1}\sin^2(\theta/2)d\theta$) one finds that
the subspace of $\bigwedge^3V_\lambda^*$ on which $H_\lambda$ acts
trivially, call it $W_\lambda$, has dimension
\beq
\label{31}
\dim\, W_\lambda=\int_{H_\lambda} \chi_0\chi\, d\mu_\lambda=
\left\{\begin{array}{cc}
8 & \lambda>0 \\
4 & \lambda=0.\end{array}\right.
\eeq

Let $\mu_1,\mu_2$ denote the invariant 1-forms $\mu_1:=\lamvec\cdot d\lamvec$
and $\mu_2:=\lamvec\cdot\sigvec$, and $\nu$ denote the invariant 2-form
$\nu:=d\lamvec\cdot\sigvec-\sigvec\cdot d\lamvec$.
Consider the following eight 3-forms:
$$
\rho_1^-=d\lambda_1\wedge d\lambda_2\wedge d\lambda_3,\quad 
\rho_2^-=\mu_1\wedge\nu,\quad
\rho_3^-=d\lamvec\cdot(\sigvec\times\sigvec),\quad
\rho_4^-=\mu_1\wedge\lamvec\cdot(\sigvec\times\sigvec),
$$
\beq
\label{33}
\rho_1^+=\sigvec\cdot(d\lamvec\times d\lamvec),\quad 
\rho_2^+=\mu_1\wedge\lamvec\cdot(d\lamvec\times\sigvec+
\sigvec\times d\lamvec),\quad
\rho_3^+=\mu_2\wedge\nu,\quad 
\rho_4^+=\sigma_1\wedge\sigma_2\wedge\sigma_3.
\eeq
The superscripts $\pm$ denote the decomposition of $\bigwedge^3\msi$ into
$\pm 1$ eigenspaces of the pullback $\Pp^*$ of the antiholomorphic isometric
involution $\Pp:M\mapsto \ol{M}$ (entrywise complex conjugation). Clearly
$\rho_a^\pm$ are by construction $SO(3)\times SO(3)$ invariant. Furthermore,
their restrictions $\rho_a^\pm|_{M_\lambda}$ span $W_\lambda$ for all
$\lambda\geq 0$
by (\ref{31}). Hence any invariant 3-form $\psi$ may be decomposed as
\beq
\label{A}
\psi=\sum_{a=1}^4(C_a^-\rho_a^-+C_a^+\rho_a^+)
\eeq
where $C_a^\pm$ are functions of $\lambda$ only. 

We now construct the Hodge isomorphism restricted to such invariant forms.
It suffices to construct $*:W_\lambda\ra W_\lambda$. Note that
$\Pp^*\vol=-\vol$ ($\vol$ being the volume form) and by definition
\beq
*\psi\wedge\omega=\langle\psi,\omega\rangle_\gamma\vol
\eeq
for all $\omega$, so $*$ must have the block structure
\beq
*=\left[\begin{array}{cc}
0 & *_+ \\ *_- & 0 \end{array}\right]
\eeq
where $*_\pm:W_\lambda^\pm\ra W_\lambda^\mp$ and $W_\lambda=W_\lambda^-
\oplus W_\lambda^+$ again denotes the eigenspace decomposition with
respect to $\Pp^*$. Straightforward linear algebra establishes that
\bea
*_+&=&\frac{1}{\Lambda^3}\left[\begin{array}{cccc}
-2\lambda^2 & 0 & 8\lambda^2 & -8 \\
-(\Lambda^2+\lambda^2) & 0 & 2\Lambda^2+4\lambda^2 & -4 \\
-\Lambda^4/2 & -\lambda^2\Lambda^4/2 & 0 & 0 \\
-(\Lambda^2+\lambda^2/2) & \Lambda^4/2 & 
2(\Lambda^2+\lambda^2) & -2
\end{array}\right],\nonumber \\
*_-&=&\frac{1}{\Lambda}\left[\begin{array}{cccc}
\lambda^2/2 & -2\lambda^2 & 2 & 2\lambda^2 \\
-1/2 & 2 & 0 & -2 \\
(\Lambda^2+\lambda^2)/4 & (\Lambda^2+2\lambda^2)/{2} & 1 &
\lambda^2 \\
(\Lambda^2+\lambda^2)^2/{8} & 
-\lambda^2(\Lambda^2+\lambda^2)/{2} &
\lambda^2/{2} &
\lambda^4/{2} \end{array}\right]
\eea
relative to the bases (\ref{33}). Note that, as claimed above, $*$ is
independent of $A$, and that $*_-*_+=*_+*_-=-\I_4$ so that $**=-\id$, as of
course it must. 

To obtain a harmonic 3-form $\psi$, it is sufficient (but not necessary,
$\msi$ being noncompact) to demand that $\psi$ be closed and co-closed
($d\psi=d*\psi=0$).  Using the notation of (\ref{A}),
\bea
d\psi&=&(C_2^-+\frac{\dot{C}_3^-}{\lambda}-C_4^-)\mu_1\wedge\rho_3^-+
(\frac{\dot{C}_1^+}{\lambda}-2C_2^+)\mu_1\wedge\rho_1^++
(\frac{\dot{C}_3^+}{\lambda}-C_2^+)\mu_1\wedge\rho_3^+ \nonumber \\
&& 
+\frac{\dot{C}_4^+}{\lambda}\mu_1\wedge\rho_4^++
(C_3^+-\frac{C_1^+}{2})\nu\wedge\nu
\eea
where $\dot{C}_a^\pm:=dC_a^\pm/d\lambda$. So $\psi$ is closed if and only
if
\bea
\label{1}
&& C_2^-+\frac{\dot{C}_3^-}{\lambda}-C_4^-=0 \\
\label{2}
&&
C_1^+=2C_3^+,\quad
\dot{C}_3^+=\lambda C_2^+, \quad
\dot{C}_4^+=0.
\eea
Clearly $\psi$ is closed and co-closed if and only if $*\psi$ is closed
and co-closed. Furthermore, by the block structure of $*$, if $\Pp^*\psi=
\psi$ then $\Pp^**\psi=-*\psi$, so it suffices to find all closed and
co-closed $\Pp^*$-even 3-forms, $\psi^+=\sum C_a^+\rho_a^+$. Applying
(\ref{2}) to the coefficients of $\psi^+$, then (\ref{1}) to the coefficients
of $*\psi^+$, one obtains a system of ODEs for
$C_a^+$, the most general solution
to which, assuming nonsingularity on the exceptional orbit $\lambda=0$, is
\beq
\psi^+=at_1+bt_2
\eeq
where $a,b\in\R$ are constants, and
\bea
t_1&=&y(\lambda)\rho_1^++\frac{y'(\lambda)}{2\lambda}\rho_2^++\frac{1}{2}
y(\lambda)\rho_3^+,\quad 
y(\lambda)=2+\frac{1}{\lambda^2}\left[1
-\frac{\log(\lambda+\Lambda)}{\lambda\Lambda}\right]
\nonumber \\
t_2&=&-4\rho_1^+-2\rho_3^++\rho_4^+.
\eea
The most general $\Pp^*$-odd closed, co-closed invariant 3-form is then
\beq
\psi^-=a'*t_1+b'*t_2.
\eeq
So the whole family is 4-dimensional, and presumably coincides with that
constructed in \cite{cvegib}. Given any form in the complexification of
this family (i.e.\ allowing $a,b,a',b'\in\C$) one may construct a complex
(anti-) self dual harmonic form
\beq
\wt{\psi}=\psi\mp i*\psi,\qquad *\wt{\psi}=\pm i\wt{\psi}
\eeq
as required for a D3 brane. 

We reiterate that these forms are harmonic with respect to {\em any}
$SO(3)\times SO(3)$ invariant K\"ahler metric on $\msi$, including 
$\gamma_{L^2}$. Of particular interest in this context, as explained above,
are antiholomorphic forms. Since $\msi$ supports a Calabi-Yau metric, the
canonical bundle $\bigwedge^{(3,0)}\msi$ is trivial. Let us introduce
the invariant holomorphic 1-forms
\beq
\epsilon_a:=\sigma_a-iJ^*\sigma_a
\eeq
where $J^*$ is the pullback of the almost complex structure
$J:T_M\msi\ra T_M\msi$ as found in \cite{spe2}
(in matrix terms, $J^*=J^T$). Explicitly, one sees that at $M_\lambda$,
\beq
\epsilon_1=\sigma_1-\frac{i}{\Lambda}(2d\lambda_1-\lambda\sigma_2),\quad
\epsilon_1=\sigma_2-\frac{i}{\Lambda}(2d\lambda_2+\lambda\sigma_1),\quad
\epsilon_3=\sigma_3-\frac{2i}{\Lambda}d\lambda_3,
\eeq
for example. The canonical bundle
has nonvanishing global section $\epsilon_1\wedge\epsilon_2\wedge\epsilon_3$.
Comparing this at $M_\lambda$ with $t_2|_{M_\lambda}$ and appealing to
$SO(3)\times SO(3)$ invariance shows that
\beq
\epsilon_1\wedge\epsilon_2\wedge\epsilon_3=\frac{1}{\Lambda^2}(t_2-i*t_2).
\eeq
So the anti-self dual 3 form $t_2+i*t_2$ is harmonic and antiholomorphic.
As remarked in \cite{cvegib}, $\psi$ is not $L^2$ normalizable with
respect to the Ricci-flat metric $\gamma$. Since $*$ is independent of $A$
for invariant 3-forms, so is
\beq
||\psi||^2=\int_{\msi} *\psi\wedge\psi,
\eeq
and it follows that $\psi$ is not $L^2$ normalizable with respect to
$\gamma_{L^2}$ either. This makes its interpretation as a quantum state
problematic (as it does the D3 brane interpretation also). Does it make sense
for a non-normalizable state to be BPS? Can one still interpret 
${\cal P}(\lambda)$, where $*\psi\wedge\psi={\cal P}(\lambda)\vol$, as a
probability density?

If not then we have actually proved that there is no invariant BPS 3-form
state, i.e.\ none with maximally excited fermionic zero modes. To see this
note that such a state would have to be $L^2$ finite, invariant
 and harmonic, which
implies, by self adjointness of $d^\dagger d+dd^\dagger$ on $L^2$, that the
state is closed and co-closed, irrespective of the noncompactness of
$\msi$. But we have classified all such forms, and none is $L^2$ finite.
It would be interesting to perform a similar analysis for invariant
0-,1- and 2-forms. Here the results {\em will} depend on $A$, but seem to do so in 
quite a simple way so that again, translation between different 
$SO(3)\times SO(3)$ invariant K\"ahler geometries should be 
quite straightforward. 

\section{Special Lagrangian submanifolds}
\label{speci}

The moduli space
picture can give considerable geometric insight. For example, consider
those holomorphic maps $\phi:S^2\ra S^2$ which factor through the quotients
in
\beq
\label{m80}
\begin{CD}
S^2 @>{\phi}>> S^2 \\
@V{\pi}VV        @VV{\pi}V \\
\RP^2 @>{\wt{\phi}}>> \RP^2
\end{CD}
\eeq
where $\pi$ denotes the covering projection. The projected maps
$\wt{\phi}$ so generated are precisely the harmonic maps $\RP^2\ra\RP^2$
\cite{eellem}, or in physics language, static $\RP^2$ lumps on $\RP^2$. 
Now $\phi\in\msn=\holo_n(S^2,S^2)$ factors if and only if it 
is a fixed point of $\Pi:\msn\ra\msn$,
\beq
\Pi:\phi\mapsto p\circ\phi\circ p
\eeq
where $p:S^2\ra S^2$ is the antipodal map. In fact $\Pi$ is an 
antiholomorphic isometry of any invariant K\"ahler metric $\hat{\gamma}$, 
and the fixed point set of $\Pi$, call it $\msnt$,
 has exactly half the (real) 
dimension of $\msn$. 
Since $\Pi$ is an isometry, $\msnt\subset\msn$ is totally geodesic, and 
hence minimal. Further,
since $\Pi$ is antiholomorphic, the restriction of the K\"ahler form
$\Omega=\hat{\gamma}(J\cdot,\cdot)$ to $\msnt$ vanishes identically on
$T\msnt$ \cite{spe2}. So $\msnt$ is a minimal
Lagrangian submanifold of $\msn$. In the case of $\msi$, $\msit$ is
precisely the exceptional orbit $\lambda=0$ (diffeomorphic to $SO(3)$),
the ``bolt'' in the language of \cite{cvegib}. Choosing 
$\hat{\gamma}=\gamma$, the Ricci-flat metric, one has a connected,
embedded, minimal Lagrangian
submanifold of a Calabi-Yau manifold, which must therefore be special
Lagrangian, by a straightforward extension \cite{mok} of a theorem of
Harvey and Lawson \cite{harlaw} concerning such submanifolds in $\C^n$.  
In this picture, the special Lagrangian bolt is
viewed as the space of (twisted degree 1) harmonic maps $\RP^2\ra\RP^2$
sitting inside the space of (degree 1) harmonic maps $S^2\ra S^2$.

Cvetic et al arrive at the same conclusion (the bolt is special
Lagrangian) but via explicit consideration of the calibrated geometry of the
holomorphic harmonic 3-form $t_2-i*t_2$. The argument above gives an
alternative derivation, from a somewhat different geometric viewpoint.
Note that its key ingredient is the antiholomorphic
isometry $\Pi$ whose fixed point set is 3-dimensional. Another such isometry
is the reflexion $\Pp$ already defined, $\Pp:M\mapsto\ol{M}$. Its fixed
point set, diffeomorphic to $S^1\times \C$ \cite{spe1}, is another special
Lagrangian submanifold of $(\msi,\gamma)$. In terms of the canonical open
embedding $\msi\subset\CP^3$, it is simply the intersection of $\msi$ with
the obvious totally real subspace $\RP^3\subset\CP^3$.

\section*{Acknowledgments}

The author wishes to thank Mark Haskins for useful discussions. He is an
EPSRC Postdoctoral Research Fellow in Mathematics.

\end{document}